\documentclass[]{article}

\usepackage{blindtext} 

\usepackage[sc]{mathpazo} 
\usepackage[T1]{fontenc} 
\usepackage{microtype} 

\usepackage[english]{babel} 

\usepackage[hmarginratio=1:1,top=32mm,columnsep=20pt]{geometry} 
\usepackage[hang, small,labelfont=bf,up,textfont=it,up]{caption} 
\usepackage{booktabs} 

\usepackage{lettrine} 

\usepackage{enumitem} 
\setlist[itemize]{noitemsep} 

\usepackage{abstract} 

\usepackage{titlesec} 
\renewcommand\thesection{\Roman{section}} 
\renewcommand\thesubsection{\roman{subsection}} 
\titleformat{\section}[block]{\large\scshape\centering}{\thesection.}{1em}{} 
\titleformat{\subsection}[block]{\large}{\thesubsection.}{1em}{} 

\usepackage{fancyhdr} 
\usepackage{titling} 

\usepackage{hyperref} 

\usepackage{graphicx,amssymb,mathrsfs,amsmath,pifont,amscd,latexsym,color,fancyhdr,CJK,url,hyperref,multirow, subfigure, colortbl,tabularx,threeparttable, booktabs,balance,natbib}
\usepackage{flushend}

\pretitle{\begin{center}\Huge\bfseries} 
\posttitle{\end{center}} 
\title{Towards Release Strategy Optimization for Apps \\in Google Play} 
\author{%
\textsc{Sheng Shen, Xuan Lu, Ziniu Hu}
\\[1ex] 
\normalsize Peking University \\ 
\normalsize \href{mailto:luxuan@pku.edu.cn}{\{sincera, luxuan, bull\}@pku.edu.cn} 
}
\date{\today} 
\renewcommand{\maketitlehookd}{%
\begin{abstract}
In the appstore-centric ecosystem, app developers have an urgent requirement to optimize their release strategy to maximize the success opportunity of their apps. To address this problem, we introduce an approach to assisting developers to select the proper release opportunity based on the purpose of the update and current condition of the app. Before that, we propose the interval of an update to its previous update to characterize release patterns, and find significance of the release opportunity through empirical analysis. We mined the update-history data of 17,820 apps from 33 categories in Google Play, over a period of 105 days. With 41,028 releases identified from these apps, we reveal important characteristics of update intervals and how these factors can influence update effects. We suggest developers to synthetically consider app ranking, rating trend, and what to update in addition to the opportunity before releasing an app version. We propose a Multinomial Naive Bayes model to help decide an optimal release opportunity to gain better user adoption.
\end{abstract}
}

\begin{document}

\maketitle

\section{Introduction}
In the past few years, the ``app economy'' has been growing fast and gaining revenues to developers. As of 2015, millions of apps have been developed and distributed in app marketplaces such as Apple Store and Google Play. It was reported that the yearly revenue from the Apple AppStore has reached over 20 billion USD as of 2015,\footnote{http://www.computerworld.com/article/3019716/apple-ios/apples-cut-of-2015-app-store-revenue-tops-6b.html} and the aggregated revenue from all app stores is expected to reach 70 billion by 2017.\footnote{http://www.digi-capital.com/reports} The great profits and rapidly released apps make the app marketplaces very competitive, and developers have an urgent requirement to distinguish their own apps out of various competitors that provide similar features. 

Release (or update) strategies are believed to be important to app success. \cite{nayebi2016mobile} found from their survey that experienced developers believe that release strategy affects user feedback. \cite{Martin:FSE16} demonstrated that releases can have causal impact to user feedback in Google Play. \cite{mcilroy2016fresh} suggested that developers should not shy away from updating their apps very frequently. However, the rationale behind the relationship between app success and release strategy is still not comprehensively studied. Hence, directive information is lacked for developers to optimize their release strategy.

In this paper we propose an empirical study to help developers decide the opportunity of release to maximize positive feedback from users at scale. Our approach is conducted with 17,820 apps from Google Play. Compared to previous study (\cite{mcilroy2016fresh}), we have some new insights as follows.

First, we find that the release frequency is not always a comprehensive indicator for app release strategy. For example, an app is updated 4 times in a week but stays stable in the following 3 weeks. Another app is updated once for each of the 4 weeks. Both of the two apps have a weekly release frequency. However, the update results of the two apps can be different, because users are concerned about and even annoyed by the frequent updates\footnote{http://www.makeuseof.com/answers/many-android-apps-update-frequently/} (i.e., the 4 updates in one week). To alleviate this problem, we propose to measure the frequency of each update with the \textit{interval} information against its previous update. The \textit{interval} can characterize the locality of releases instead of the average frequency. We divide the update \textit{interval} to three groups of \textit{successive}, \textit{normal}, and \textit{sparse} and find differences in feedback.

Second, the effect of a release should not only be related to the release behavior itself, but also an aggregated result of the quality (\cite{Lu:2017ubicomp}) and accumulated user adoption of the corresponding app. Intuitively, a user may be more tolerant to successive updates for the app that she likes than that she dislikes. We employ the \textit{ranking} of an app, which is provided by Google Play, to indicate the position of the app in its category. Analysis results show that the change of ranking after a successive update suggests the \textbf{Matthew Effect}, i.e., \textit{the high-ranked apps get higher ratings and the lowly-ranked apps get lower ratings}. Additionally, we leverage the \textit{rating trend}, which is detected from the rating curve, to indicate the current situation of an app. We find that a timely subsequent release can take a turn for a descending rating trend with a higher probability than a late release. 

Third, user feedback can be different to releases with different purposes. Traditional reasons for app releases include bug fixes, new features, and library updates (\cite{Gui2015Truth}), which also apply to traditional software releases. With the unique and low-cost deployment platform such as Google Play in the mobile era, new version of apps are also released for additional reasons, such as stimulating  downloads and ratings (\cite{Comino2015Updates}). Terms indicating different purposes appear unequally in releases with positive and negative impacts (\cite{Martin:FSE16}). We move a step further to see how the purpose affects feedback with different update frequency. Results imply that developers should not shy away from fixing bugs by performing hotfixes, but be cautious not to change features too frequently.

We follow up on the empirical results with a Multinomial Naive Bayes model to predict update effect based on the aforementioned factors. The prediction achieves a high accuracy of 76\% in a 5-fold cross validation. By maximizing the effect variable, we provide developers with an optimal \textit{interval} for the next update. Developers can refer to the suggested \textit{interval} to optimize their release plan and optimize development process.

More specifically, we make the following main contributions:

\begin{itemize}
\item We propose to measure app release patterns in finer grain and use \textit{interval} to characterize the local features of releases. Various findings and implications are derived from the observations.

\item We conduct an empirical analysis to explore the relation among update effect and \textit{app ranking}, \textit{rating trend}, and \textit{update purpose} using large scale of apps from Google Play, i.e., 41,028 releases of 17,820 apps.

\item We propose the first approach to optimizing app release strategy by training a Multinomial Naive Bayes model and give actionable suggestions to Android app developers to optimize their release strategy.
\end{itemize}

The rest of the paper is organized as follows. Section~\ref{definition} introduces the concepts and terms used in this paper. Section~\ref{rqs} explains the research questions. Section~\ref{dataset} describes the data set. Section~\ref{update}, \ref{feedback}, and \ref{timing} study the three research questions posed in Section~\ref{rqs}, respectively. Section~\ref{discussion} discusses threats to validity of this study. Section~\ref{related} describes the related work of app release. Section~\ref{conclusion} concludes the paper.

\section{Definitions}\label{definition}
In this section we define the terms and metrics used in this paper.

\subsection{Update Interval}\label{interval}
Instead of using average frequency to represent release pattern of an app (\cite{mcilroy2016fresh}), in this study we focus on each single update of apps. That is, for each update, we count the days of \textit{interval} to its previous one as the \textit{frequency} of this update. For example, let $u_{i}$ denote the $i$th update of app $\mathbb{A}$. $u_{i}$ was conducted on the day $t$, and the update $u_{i+1}$ of $\mathbb{A}$ was conducted on the day $t+k$. The interval of $u_{i+1}$ to $u_{i}$ is $k$. In this way, each update is described with the \textit{interval}, and the releases of an app can be represented as a sequence of intervals.

To investigate the difference of user feedback in updates with different intervals, we first categorize the updates according to their intervals into three groups, i.e., \textit{successive updates}, \textit{normal updates}, and \textit{sparse updates}.

\noindent $\bullet$ \textbf{Successive Updates} are updates that have intervals between 1 day to 5 days to its previous update. Users concern about the reason and necessity of an additional update right after an existing one. Additionally, app quality can be unstable with fast iteration. We assume it interesting to see the change of user feedbacks to such updates. 

\noindent $\bullet$ \textbf{Normal Updates} are updates that have intervals between 5 days to 20 days to its previous update. Such updates can be considered to perform routine maintenance and improvement of apps. 

\noindent $\bullet$ \textbf{Sparse Updates} are updates that have intervals longer than 20 days to its previous update. Such intervals are quite long in the fast iterative environment. A sparse update can be a release of version with new features or just a bugfix. We would like to see the influence of sparse updates.

\subsection{Rating}
Rating is an important indicator of popularity and success of an app. For different versions of apps, ratings rise or fall to capture the satisfaction levels of users (\cite{ruiz2016examining}). We adopt rating to evaluate user feedbacks to updates in this study.

In addition to reacting to updates, ratings of apps can be influenced by multiple factors including latent ones. In other words, ratings are unstable (\cite{Martin:FSE16}). In comparison, a fitting curve can describe user adoption to both the app itself and recent changes with disturbance omitted, which is more reliable. We fit the trend of rating from the curve and detect \textit{turning points} after which the trend changes significantly. The approach to detecting the turning points is described in Section~\ref{rqs}.

\subsection{Factors Influential to Update Effects}
Effect of an update to an app is related to various kinds of factors. One important factor is the opportunity of releasing, i.e., the \textit{interval} of the update. In addition, the current popularity of the app which is measured with its ranking, the rating trend which reflects user adoption of the app in the dynamic perspective, and the purpose of update (i.e., fixing bugs or adding new features) can also contribute to update effects.
  
\noindent $\bullet$ \textbf{Ranking.} Google Play provides a ranking list for top apps in each category. Each list contains 540 apps. The ranking of each app in this list reflects aggregated popularity of the app in relative to other apps in the same category. 
Although the algorithm of ranking works as a blackbox, user adoption is considered to be involved in the algorithm. 
We assume that users hold different levels of tolerance to highly ranked apps and lowly ranked apps.
The lists keep updating and we crawl them daily.

\noindent $\bullet$ \textbf{Rating Trend.} Due to various reasons, user adoption of an app can change. For example, ratings of a famous app can decline because of unstable networking service. Rating of a little-known app can increase for its fascinating new functions. We wonder whether the current trend of rating show inertia when coming with a new update. In other words, will a new update assist a good trend or reverse a bad trend?

\noindent $\bullet$ \textbf{Update Purpose.} Existing research has revealed that bug fixing occupies a large proportion of development efforts. In addition, terms of `bug', `fix', `new', and `feature' are more common in releases that positively affected ratings, as opposed to those that negatively affected ratings (\cite{Martin:FSE16}). In this paper, we take a further step to reveal the reasons of both positive and negative effects of an update with either purpose of fixing bugs or adding new features.

\section{Research Questions}\label{rqs}
In this section we propose three research questions and explain the methodology to answer them.

\noindent \textbf{RQ1: Is \textit{interval} of update a good metric to characterize app releases?}

Before studying user feedback of releases, we first conduct an empirical analysis about the distribution of update intervals in Google Play in Section~\ref{update}. Since we assume that the metric \textit{interval} is better in characterizing local release patterns than average frequency, we determine whether successive updates are significant in all the updates, and whether it is common in apps. If the catholicity of successive updates is supported, this motivates further analysis of them. 

Furthermore, we explore the philosophy behind different patterns of intervals. It is expected that the features of update interval can reveal more information about the developers such as the attitudes towards release strategy.

\noindent \textbf{RQ2: What impacts update feedbacks?}

We measure update effects using the app rating as user feedback. Because of the instability of app ratings (\cite{Martin:FSE16}), it is hard to trace the rating change back to release impacts. Hence, we focus on \textbf{\textit{significant updates}} within \textit{time-lag} slots before \textit{turning points}, and explore the influential factors in Section~\ref{feedback}. 

\noindent $\bullet$ \textbf{Turning Points}. 
A turning point is where the rating trend change significantly in the fitted rating curve. We detect the turning point in two steps. First, the ratings of a specific app could be aggregated into a time series vector $T$. In this way the detection problem can be framed into a segmentation problem, which can be traditionally settled with producing the best representation of $T$ using only $K$ segments (\cite{sugiura1994testing}, \cite{ge2001segmental}). Second, we employ \textit{Top-Down online algorithm} to train a concurrent Piecewise Linear Representation (\cite{chua1986canonical}) model (i.e., PMR model) on $T$. For clarity, this method works by considering every possible partitioning of the time series vector $T$ and splitting it at the best location. Both subsections are then tested to see if their approximation error is below the user-specified threshold, which we set as the median consecutive changes of $T$. Otherwise, the algorithm recursively continues to split the sub-sequences until all the segments have approximation errors below the threshold.

\begin{figure}[hbt]
\centering
  \includegraphics[width=0.5\textwidth]{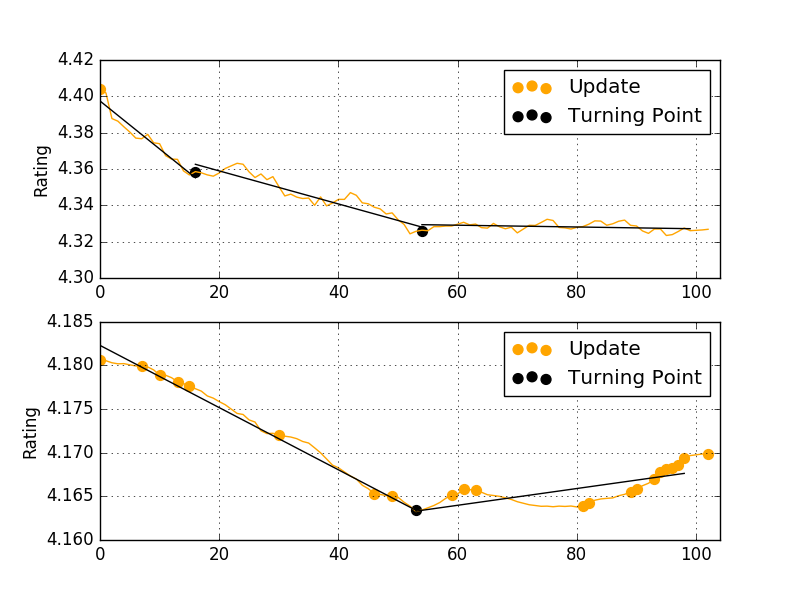}
  \caption{Turning Points}
  \label{fig:tp}
\end{figure}

Obviously, various models besides PLR model can support the segmentation mining, such as the Bayesian Structural Time-Series model (\cite{wilkinson1999statistical}) applied by (\cite{Martin:FSE16}). This method focuses on the probability that model generate by comparing an occurred time series vector post-release with the counterfactual prediction. A low $p$-value ($p \leq 0.01$) will indicate a significant change of rating, which means a turning point has been detected. However, we choose to fit the trend of time series vector by Linear Interpolation (\cite{de1978practical}), and the turning point occurs when the trend changes. 
The way of detecting turning points is demonstrated in Figure~\ref{fig:tp}.

\noindent $\bullet$ \textbf{Time Lag.} Given the fact that users need time to react to a new version of apps (to update the app on their devices, to get familiar with new functions, etc.) by rating, the lag-effect between update curve and rating curve undoubtedly exists. To detect such effect, we first model update points and ratings into two series vectors, $T1$ and $T2$, respectively. Smooth operations are employed on $T1$ and $T2$ to eliminate the noise. Second, we use the Pearson product-moment correlation coefficient (i.e., PPCC) (\cite{lee1988thirteen}) to estimate how update and rating are correlated with different time lags. Similar to the correlation study between the proposed features of a company and the financial time series for its stock (\cite{ruiz2012correlating}), we apply this time lag detection method to each app and calculate the PPCC value for each pair of vectors with different time lags. A low $p$-value ($p \leq 0.01$) indicates that the update vector correlates with the rating vector significantly. 

By accumulating the percentage of significantly correlated vectors of each app with different time lags, it is found that 4 days lag between update vector and rating vector is the largest as {48\%} where the values at 0 day lag is {35\%}. The analysis suggests the existence of time lag between update behavior and rating reaction, which in turn indicates that an update within a proper time slot before a rating change is correlated with the change with high possibility.

\noindent \textbf{RQ3: How to optimize the strategy of app release?}

It is a complex task to optimize release strategy because multiple factors, even latent ones, can influence the effect of updates. Based on the factors studied in this paper, we take the first step to help developers to select an optimized time to release next version of  their apps through a Multinomial Naive Bayes model in Section~\ref{timing}. 

\section{Data Set}\label{dataset}
We collected a set of 17,820 apps of 33 categories in Google Play from November 25, 2016 to March 10, 2017 (105 days in total). The gathered apps all appeared in the `top 540' lists of each category during this period. We adopt this list because the included apps are likely to have many releases, received feedbacks from users, and have reliable data of ratings, yet alleviate the App Sampling Problem (\cite{martin2015app}).

The crawler cycles over each of the 17,820 apps in the span of 24 hours. Finally, 41,028 updates are collected. We then extract the information we need in our study, i.e., ranking, rating, text in \textit{what's new}, and version of each app. 

\begin{figure}[hbt]
\centering
  \includegraphics[width=0.5\textwidth]{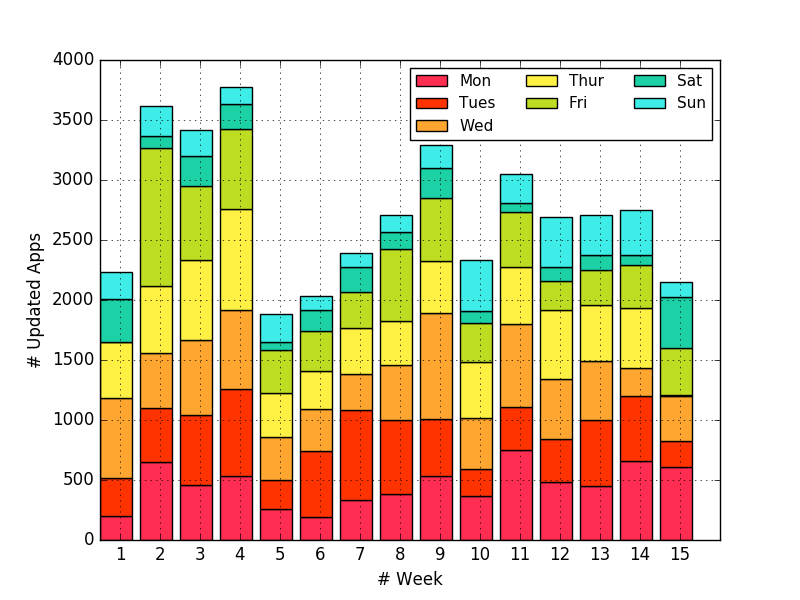}
  \caption{Distribution of App Updates}
  \label{fig:distribution_u}
\end{figure}

We take a first look at the data set by plotting the distribution of updates in each week. As shown in Figure~\ref{fig:distribution_n}, each bar represents a week and each part of the bar denotes number of updated apps in a day. The number of updated apps varies a lot across weeks and no obvious pattern is observed. However, within each week, we find that \textbf{more apps choose to update on weekdays than weekends}. Meanwhile, \textbf{there are still apps choose to update on weekends} in some weeks. Such observations consist with developers' working mode. No significantly favored day is observed for developers to update their app. We infer that developers prefer releasing apps once they are ready to following a weekly schedule.

\section{Update Patterns}\label{update}
This section studies the first research question, i.e., \textit{Is interval of update a good metric to characterize app releases?}

\subsection{Interval Distribution}
In previous work, frequency is an average value calculated as the number of updates divided by number of weeks. We propose that the \textit{interval} of update contains more information about the local characteristics of release behaviors. In this study, the \textit{frequency} of each update is measured by the \textit{interval}.

\begin{figure}[hbt]
\centering
  \includegraphics[width=0.5\textwidth]{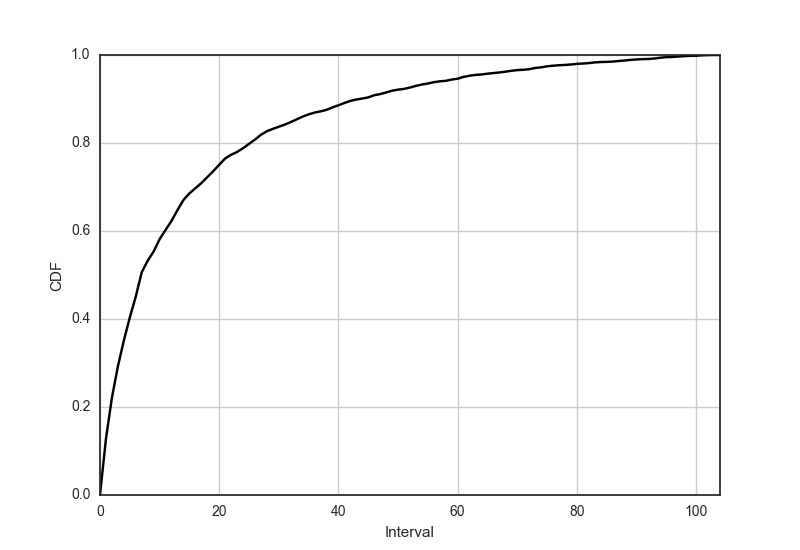}
  \caption{CDF of Update Interval}
  \label{fig:frequency}
\end{figure}

We plot the cumulative distribution function (CDF) of updates with different intervals in Figure~\ref{fig:frequency}. The intervals spans largely from 1 day to over 100 days, which implies the necessity of considering more metrics than average frequency. The distribution shows that 75\% percent of updates have an interval of no more than 20 days. In addition, 40\% of updates have an interval between 1 day to 5 days. The high proportion of successive proves that \textbf{successive updates constitute a large part of all updates} and motivates us to study them further. The distribution of 40\%  successive updates, 35\% normal updates, and 35\% sparse updates supports the reasonability of the categorization of update intervals in Section~\ref{interval}.

\begin{figure}[hbt]
\centering
  \includegraphics[width=0.5\textwidth]{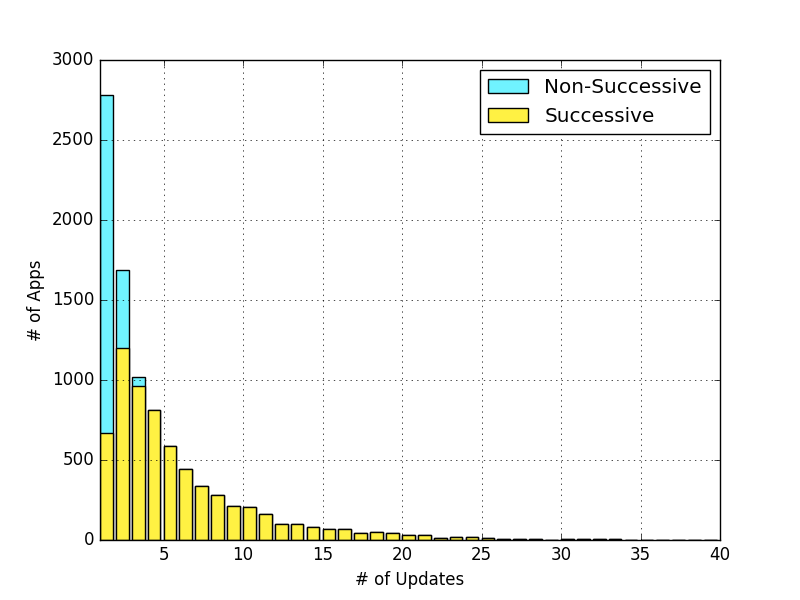}
  \caption{Distribution of Apps}
  \label{fig:distribution_n}
\end{figure}

An important concern is the catholicity of successive updates in apps, given that they hold a high proportion in all updates. We present the app distribution to the number of updates it experienced in our data set. Considering that 54\% of apps were not updated, we only use the 46\% of apps that experienced at least one update. As Figure~\ref{fig:distribution_n} shows, the number of apps sharply decreases when the number of updates increases. Such results implies that most apps do not update, and most of the updated apps only update for only a few times during 15 weeks.

Furthermore, approximately 56\% of the updated apps experienced the successive-update pattern for at least one time, which means the update interval is less than 5 days, as Figure~\ref{fig:distribution_n} demonstrates. The proportion is quite high and proves that \textbf{successive updates occur in the updates of most apps}. Additionally, a sizable subset (˜{15\%} of the updated apps) experienced the normal-update pattern, which means an update interval less than 21 days.

Based on the general analysis, we then investigate the interval distribution of apps in different categories. By calculating the possibility of successive updates in each category, we find that apps in \textit{tools} are most likely to experience successive updates, followed by apps in \textit{social} category. Apps in categories such as \textit{art} and \textit{medical} are less likely to have successive updates.

\noindent \fbox{%
\parbox{1\textwidth}{
     \textbf{Finding (F1)}: Successive updates constitute 40\% of all updates in our data set, and occur in approximately 56\% of the updated apps. The catholicity of successive updates motivates further study of them.} } 

\subsection{Update Consistency}
As we proposed, the update interval can reflect local characteristics of release behaviors. We are then curious if the update interval pattern can reveal the philosophy of developers. 

\begin{figure}[hbt]
\centering
  \includegraphics[width=0.5\textwidth]{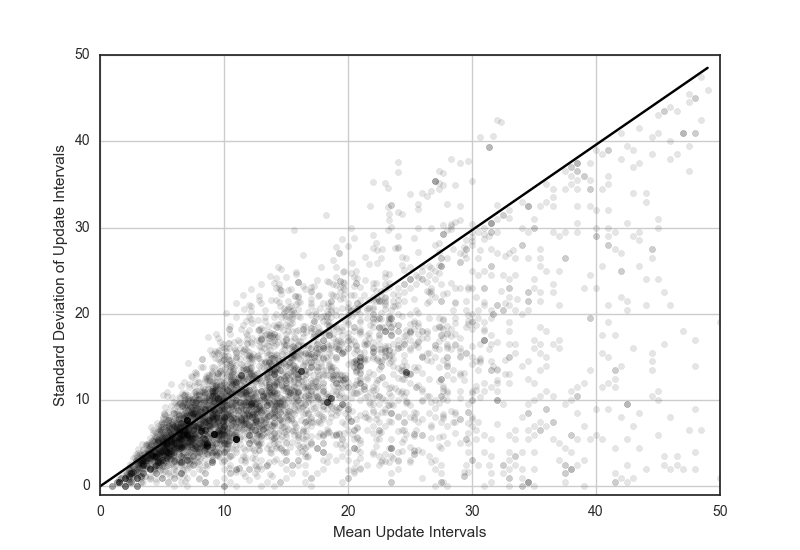}
  \caption{Standard Deviation and Mean Value of Update Intervals of Apps}
  \label{fig:distribution_v}
\end{figure}

To this end, we calculate the mean value and the standard deviation of update intervals for each app. The relation between the two variables is shown in Figure~\ref{fig:distribution_v}. As is observed, apps with lower mean update interval tend to have lower standard deviation of update intervals. We apply a linear regression model and find it fits well with the R-square as 0.77 and $p$-value $\leq0.001$. 

A high mean value of an app represents high possibility of having sparse updates, and a low mean value represents high possibility of having successive updates. The standard deviation indicates the possibility and extent of changing update intervals. A considerable quantity of apps are possible to have a relatively consistent release schedule as they are clustered in the down-left part of the figure. Hence, the linear relation between the two variables suggest that \textbf{app developers tend to perform consistent release strategy (i.e., similar update intervals) if they release app more successively}. Similarly, developers who release apps sparsely tend to choose release opportunity with quite different update intervals.
  
\noindent \fbox{%
\parbox{1\textwidth}{
     \textbf{Finding (F2)}: Characteristics of update intervals can reflect developers' philosophy in releasing apps. Developers favoring successive updates tend to perform releases routinely, while those favoring sparse updates are more likely to choose a random opportunity to release their apps.} } 

\subsection{Local Pattern Clustering}
Given that considerable apps tend to release new versions consistently, we wonder whether update behaviors show some patterns in a period. More specifically, we extract short vectors with $t$ elements from release sequences for each app, and cluster such vectors to find potential patterns. 

For each update in a release sequence, we model update behaviors in the following $t$ days using a time series vector. Each element in the vector is valued as $1$ if there is an update on the corresponding day, $0$ otherwise. We then employ two standard algorithms, the t-SNE model and the Ward hierarchical method, to perform the clustering task.

The \textbf{t-SNE model} stands for t-distributed stochastic neighbor embedding algorithm (\cite{maaten2008visualizing}), a nonlinear dimensionality reduction technique. This model is particularly well-suited for embedding high-dimensional data into a space of two or three dimensions. Specifically, it constructs a probability distribution over the high-dimensional objects in the way that similar objects have a high probability to be picked. Based on this distribution, t-SNE defines another similar probability distribution over the points in the low-dimensional map and minimizes the Kullback-Leibler divergence (\cite{kullback1951information}) between the two distributions with respect to the locations of the points in the map. In this way, t-SNE method could model each high-dimensional vector by a two- or three-dimensional point where similar objects are modeled by nearby points and dissimilar objects are modeled by distant points.

The \textbf{Ward hierarchical method} (\cite{ward1963hierarchical}) is specifically a criterion that models the cluster problem into a variance problem. Basically, this method conducts agglomerative hierarchical clustering procedure. At each step, the model chooses the pair of clusters to merge based on the minimal value of the total within-cluster variance.
  
\begin{figure}[hbt]
\centering
  \includegraphics[width=0.5\textwidth]{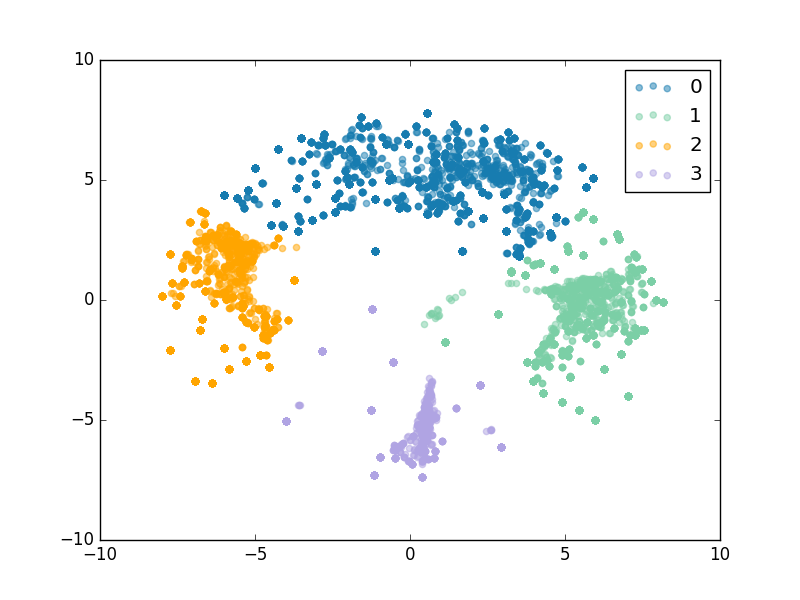}
  \caption{Clustering of Local Release Patterns}
  \label{fig:general_c}
\end{figure}

For a given sliding window of $t$ days, we first extract the time series vectors for each update. 
We choose the sliding window of 13 days (\cite{mcilroy2016fresh}) and the local pattern following each update is then modeled with such vectors. Secondly, we employ the t-SNE model to embed the space of all vectors into a two-dimensional map. In this way, the similarity between update vectors keeps with dimensionality reduced. As demonstrated in Figure~\ref{fig:general_c}, similar vectors are modeled by nearby points and dissimilar vectors are modeled by distant points. Thirdly, to find general patterns from the embedded map, we use Ward hierarchical method to classify the points into 4 clusters. Points in different clusters are colored in Figure~\ref{fig:general_c}. 

The clustering results implies that the update behaviors after an update present several patterns. This finding suggests a direction to study behaviors of developers, although some of them update their apps quite inconsistently. 

\begin{figure}[hbt]
\centering
  \includegraphics[width=0.5\textwidth]{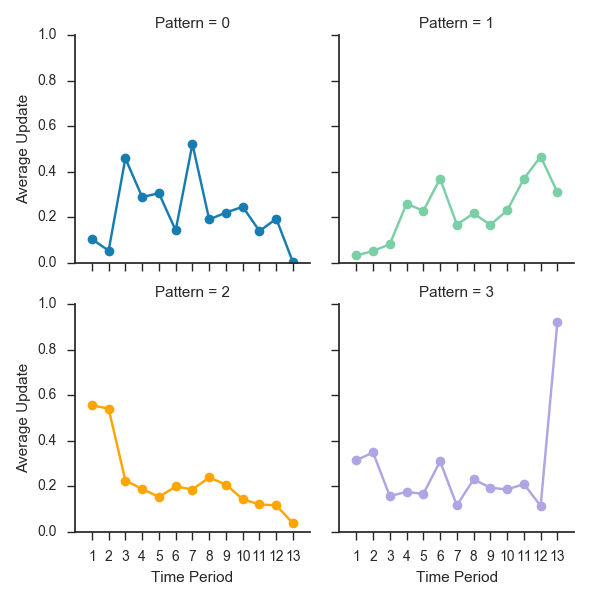}
  \caption{Aggregated Release Patterns}
  \label{fig:general_4}
\end{figure}

We additionally characterize the 4 clusters by aggregating vectors in each cluster and visualize them in Figure~\ref{fig:general_4}. The aggregated vectors, with average values of original vectors in each cluster, present 4 distinguished release patterns. Given an update happened on \textit{day 0}, the possibility of subsequent update in \textit{pattern 0} peaks on \textit{day 3} and \textit{day 7} and descends to 0 on \textit{day 13}. The possibility of update in \textit{pattern 1} follows an approximately ascending trend to \textit{day 12} with a small peak on \textit{day 6}. \textit{Pattern 2} shows high possibility of subsequent updates on \textit{day 2} and {3}, which implies successive updates, and then the possibility descends. \textit{Pattern 3} shows significant high possibility of updates on \textit{day 13}, which is a normal update in about 2 weeks, after comparatively small fluctuations.  In general, developers show different preferences to performing successive updates, weekly updates, and bi-weekly updates in these 4 different release patterns.
  
\noindent \fbox{%
\parbox{1\textwidth}{
     \textbf{Finding (F3)}: Local release patterns can be detected from release behaviors after an update. With a 13-day sliding window, developers are found to present 4 general patterns, each with different preferences to performing successive updates, weekly updates, and bi-weekly updates. } } 

\section{Update Effects}\label{feedback}
This section studies the second research question, i.e., \textit{What impacts update feedbacks?} 

The effect (or the user feedback) of an update should be influenced by multiple factors. We choose to measure user feedback with app rating. We propose three important factors that can affect ratings, i.e., the current ranking of the app, the rating trend of the app, and the purpose of the update. In the three update groups , i.e., \textit{successive updates}, \textit{normal updates}, and \textit{sparse updates}, we analyze performance of the three factors respectively.

\subsection{Ranking}
\textit{Ranking} of an app, which is evaluated by Google Play, reflects the relative position of the app in a category. Apps with higher rankings are often with high quality or popularity, while those with lower rankings can be more sensitive to external changes. We hypothesis that apps with different ranks achieve different feedbacks after updates. 

To this end, we choose to group the updates from two dimensions. The first dimension is the update interval, and the second is the direction of app reaction which is reflected on the rating. The rating of app can increase after an update, which indicates a \textit{positive} reaction. Accordingly, a \textit{negative} reaction means decrease of app rating after an update. In this way, we have 6 groups of updates, i.e., \textit{successive-positive}, \textit{successive-negative}, \textit{normal-positive}, \textit{normal-negative}, \textit{sparse-positive}, and \textit{sparse-negative}. 

\begin{figure}[hbt]
\centering
  \includegraphics[width=0.5\textwidth]{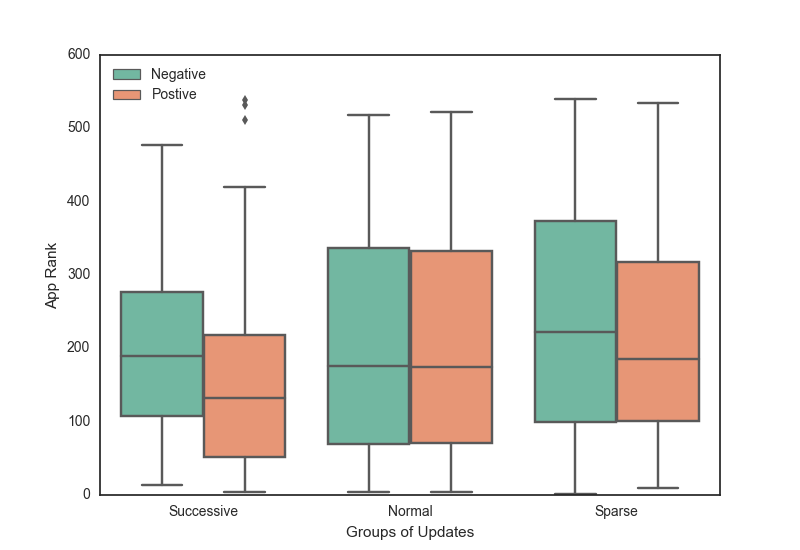}
  \caption{App Rank in Different Update Groups}
  \label{fig:rank}
\end{figure}

We then investigate the distribution of app rankings before the update in different groups, as shown in Figure~\ref{fig:rank}. Results show that the rankings are lower in the \textit{successive-positive} group (median ranking=131) than those in the \textit{successive-negative} group (median ranking=190). We perform Mann-Whitney U test between the rank distributions in these two update groups and witness significant difference ($p$-value=0.001). Such a result indicates that high-ranked apps tend to get positive feedback after successive updates in comparison to lowly-ranked apps. Meanwhile, rank distributions are not significantly different between \textit{sparse-positive} and \textit{sparse-negative} groups ($p$-value=0.291), or between \textit{normal-positive} and \textit{normal-negative} groups ($p$-value=0.323). 

The phenomenon in the ranking feedback is kind of \textbf{Matthew effect} that described in sociology, i.e., \textit{the rich get richer and the poor get poorer}. If the update is going to be successive to its previous update, apps with high rankings are more likely to get a higher rating while those with low rankings are more likely to get a lower rating. It is reasonable because the factors of people - users' attitudes, are important to the rating of apps. It is supported that users are sensitive to successive updates. Additionally, users may tend to believe that a successive  update is necessary and beneficial to a high-ranked app, while to be annoyed for a lowly-ranked app. This is effect of accumulated advantage of apps' performance and popularity in history. 

Given such results, developers should be careful before updating their apps. For high-ranked apps, developers can choose an update frequency according to their needs. But for low-ranked apps, a conservative strategy of updating normally is suggested.

\noindent \fbox{%
\parbox{1\textwidth}{
     \textbf{Finding (F4)}: App rating in response to successive updates is kind of \textbf{Matthew effect}. High-ranked apps tend to gain higher ratings while low-ranked apps gain lower ratings. Developers of apps with low rankings are suggested to maintain updating frequency not to be too frequent.}}  
     
\subsection{Rating Trend}
Although update feedback can be affected by accumulated advantage of apps, updating apps is a main way for developers to make their apps better, e.g., to stimulate downloads or ratings. We are curious whether an update can ``save'' an app by reversing or slowing down the accumulated getting-worse direction. If so, does the frequency of the updates matter to the resulted ratings?

We leverage \textit{rating trend} to represent the accumulated direction of user adoption of apps, and see what happens to a descending trend after an update. We consider only the \textbf{significant updates}, which are the nearest update within the time-lag slot before turning points. In total 1,905 updates are detected for this study. For each turning point in the rating curve, we calculate the trends of rating before and after it. The trends are quantified with the slope of the fitted curve. The rating trend can be reversed to be increasing, or be still descending after the turning point, either slower or quicker. We then group the updates according to their effects to \textit{reversed increasing}, \textit{restrained descending}, and \textit{aggravated descending}. It is found that 69\% of updates show positive effects to apps with a descending rating trend, including 49\% in \textit{reversed increasing} and 20\% in \textit{restrained descending}. Such results prove that updates can be beneficial to apps with getting-worse ratings.

\begin{figure}[hbt]
\centering
  \includegraphics[width=0.5\textwidth]{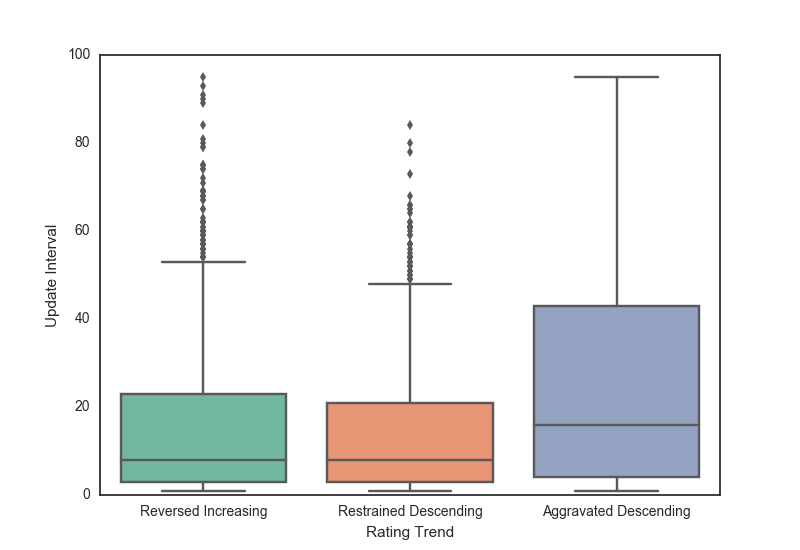}
  \caption{Update Interval in Updates Causing Different Influence after Descending Rating Trend}
  \label{fig:rating}
\end{figure}

However, there are still updates related to worsening rating trends. We then move on to see why some updates have positive effects while others have negative effects. In specific, we plot the distribution of update intervals in the groups of \textit{reversed increasing}, \textit{restrained descending}, and \textit{aggravated descending}. The median update interval in the three groups are 8, 8, and 16, respectively. As shown in Figure~\ref{fig:rating}, update intervals are shorter in the first two groups than in the \textit{aggravated descending} group. Such a result indicates that updates with shorter intervals are more likely to ``save'' an app. An example is demonstrated in Figure~\ref{fig:tp}. Two apps in Figure~\ref{fig:tp} experience similar descending trends of rating. However, with more updates, the descending trend of the second app slowly  restrains and finally turns into an ascending trend. Yet the trend in the first figure remains descending. We suggest developers not to hesitate to update their apps with a descending ranking trend. 

Additionally, we conduct the same analysis to turning points after an increasing trend. No obvious difference of interval distribution is observed among the three corresponding rating trend groups, i.e., \textit{reversed descending}, \textit{restrained increasing}, and \textit{accelerated increasing}.

\noindent \fbox{%
\parbox{1\textwidth}{
     \textbf{Finding (F5)}: Successive updates can be helpful to apps undergoing a descending rating trend.} }

\subsection{Update Purpose}
In addition to \textit{ranking} and \textit{rating trend} of the app, we assume that the purpose of an update is  important to the feedback of the update. \textit{Bug fixing} and \textit{feature adding} are proved to be the most frequent terms in such release text (\cite{Martin:FSE16}). However, the relation between update results and update purpose remains not revealed. To this end, we untangle the update feedbacks to what is updated. Considering that the frequency of updates can influence the result, we keep the updates in the three groups (i.e., \textit{successive updates}, \textit{normal updates}, and \textit{sparse updates}) in this analysis.

Apps in Google Play use the field \textit{WhatsNew} as the release text to imply the difference of the a version.  We use this field of each update to extract update purpose. Note that sometimes \textit{WhatsNew} is not revised with app updates (\cite{mcilroy2016fresh}). Such updates are discarded.

\textbf{Pre-processing}. We perform information retrieval techniques on the text. The first step is to cast the text into lower case and filter them for punctuation and stop words using Porter stemmer (\cite{porter1980algorithm}). We then employ TF.IDF (\cite{manning2008scoring}) model to embed each text with a `term vector'. The trained `term vector' contains 18,728 dimensions. Each dimension stands for the relative importance of the term to the text by calculating the term frequency inverse text frequency.

\textbf{Term Importance}. We use \textit{Lasso} (\cite{Tibshirani:1996hh}), a standard linear-regression model, to correlate the influence of the update with update purposes. In specific, we regress the rating trend difference, i.e., slope of rating trend after the update minus that before the update, to each of the `top terms'. The coefficients of variables learned by \textit{Lasso} indicate the importance of the corresponding features.

\begin{figure}[hbt]
\centering
  \includegraphics[width=0.5\textwidth]{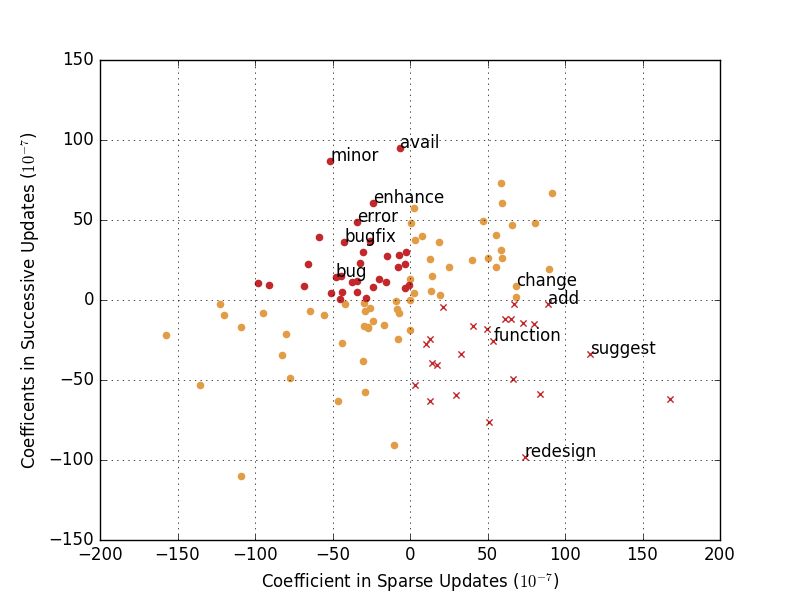}
  \caption{Update Purpose in Successive and Sparse Updates }
  \label{fig:purpose}
\end{figure}

After performing \textit{Lasso} on each group of updates, we find that the importance of terms are quite different. The terms are visualized in Figure~\ref{fig:purpose} according to their importance to updates in \textit{successive} and \textit{sparse} groups. Each point represents a term. Terms in the left part have a negative coefficient with a sparse update, while terms in the right part positive. Similarly, terms in the lower part have a negative coefficient with a successive update, while terms in the upper part positive. We are interested in the terms in the \textbf{left-upper} and \textbf{right-lower} parts. The purpose, represented by the terms, affects the successive updates and sparse updates in different directions in these two parts.

As shown in Figure~\ref{fig:purpose}, terms in the left-upper part are mostly about fixing bugs, such as \textit{bug}, \textit{bugfix}, and \textit{error}. However, terms in the right-lower part are more about changing features, such as \textit{function}, \textit{redesign}, and \textit{add}. Such results imply that influence of update purposes varies according to the frequency of update. Releasing a new version with bug fixed quickly after a version can be beneficial to an app, while changing features successively can be harmful. In contrast, changing features is more likely to be welcomed for sparsely updated apps, while fixing bugs can cause negative results.

\noindent \fbox{%
\parbox{1\textwidth}{
     \textbf{Finding (F6)}: The change of rating trend after an update is related to the purpose of the update considering the release frequency. We suggest that developers should not shy away from fixing bugs by performing hotfixes, but be cautious not to change features too frequently.} }

\section{Update Optimization}\label{timing}
Thus far, we have studied the roles of app ranking, rating trend, and purpose in updates of different frequency. Analysis results support that update results can be influenced by these aspects. The third research question then comes from the developers naturally - \textit{How to optimize the strategy of app release}? Given the complexity of factors that can influence update results, we simplify this problem by restricting factors to the three aspects we have studied. This section introduces the approach to optimizing release opportunity in two steps.

\textbf{Step 1. Modeling release effect.} We employ all the aspects discussed above into a general model to demonstrate the relation between the factors and the update effect. The model applied is Multinomial Naive Bayes (\cite{rennie2003tackling}) (i.e., the MNB model), which implements the naive Bayes algorithm for multinomially distributed data. We choose the MNB model because it is well-suited for discrete features. The factor vector derived from discrete term importance vector and other factors can be well expressed in this way.

For clarity, let $E$ denote the effect of a focal release. We model this effect as a float variable, where the positive or negative value stands for the direction of effects. Let $r$ denote the current ranking of the corresponding app at this release, which can be directly found in Google Play. Let $t$ denote the planning update interval. Let $s$, the slope of the previous fitted segmentation of rating curve detected by the Top-Down method, denote the previous rating trend. Let $C$ denote the importance vector of terms in the release text. The model is formulated as below:

	$$E_i = MultinomialNB\left (r_i, t_i, s_i, C_i\right )\: \: \:  (i\in I)$$

To validate this model, we perform a 5-fold cross validation (\cite{kohavi1995study}) to predict the direction of update effect. Accuracy of this model achieves 76\%. Considering that the update purpose is a relatively known factor to update effects (\cite{Martin:FSE16}), we simplify this model by removing $r$, $t$ and $s$ to provide a baseline. Accuracy of this baseline achieves only 54\%. We then improve the baseline by adding $r$ and $s$ and can have an accuracy of 70\%. The performance of our model with 76\% accuracy supports the advantage of considering the app ranking, rating trend, and update interval comprehensively with update purpose.

\textbf{Step 2. Optimizing release opportunity.} Based on the proposed model, the task can be conducted through an optimization procedure. Given the app ranking $r$, rating trend $s$, and importance vector $C$, we need to find a $t$ to maximize the effect $E$. 
Formally, the optimization could be expressed as:
$$t_{best} = \max_{t'\in \left \{ 0,1,\ldots,\infty\right \}}\left \{ \:MultinomialNB\left ( r_i, t', s_i, C_i\right ) \:  \right \}\: \: \:  (i\in I)$$

The output of the optimization, $t_{best}$, suggests a best opportunity to release a new version based on the current app ranking, rating trend, and the purpose of update. However, release planning is a process of trade-off among various aspects such as market feedback, app quality, and development efforts. Hence the validation of the optimization output is out of scope of this paper. The aim of this work is to reveal the rationale of update effect by well modeling it with influential factors. Developers can combine our findings and suggestions with their knowledge from experience to pursue better update effects.

\noindent \fbox{%
\parbox{1\textwidth}{
     \textbf{Finding (F7)}: Update interval with app ranking, rating trend, and update purpose can well model the effect of updates through MNB. Developers can leverage our model to optimize their release strategy.} }

\section{Threats to Validity}\label{discussion}
In this section we discuss threats to construct, conclusion, and external validity in our study.

\textbf{Construct Validity}: The analysis performed in this study are \textit{correlation} analysis but not \textit{causal} analysis. Although causal analysis can provide the most directive suggestions, it is hard to imply causality from data analysis, especially for the analysis of successive updates. However, findings from our correlation analysis are reasonable and can give useful implications to developers when managing their releases.

We use three factors (i.e., app ranking, rating trend, and update purpose) in addition to update interval to study the effects of updates. However, there will always be unknown factors that are hard to capture from data. Nonetheless, we believe the method can be applied to studying effects of new factors. 

Factors can change over time, thus the factors in the release opportunity prediction model should be functions of time variable. We assumed the factors to be relatively stable in the model. In the future we plan to model the factors with time series and improve the prediction model. 

\textbf{Conclusion Validity}: The app ranking algorithm is not released by Google Play and can change over time. If the algorithm changes, the analysis results about app ranking can not directly apply.

\textbf{External Validity}: We believe the empirical analysis and the prediction model can give actionable suggestions to developers, although the real effect can be influenced by more factors. 

The findings can not be simply applied to other app stores. Nevertheless, the methods we used in this study can be performed on data from other app stores.

\section{Related Work}\label{related}
This section looks at the literature related to update cycles of traditional software development, followed by prior work related to apps release engineering.

\subsection{Software Release} 
In traditional software development, various work has focused on enabling short, steady and consistent release cycles. For instance, \cite{de2004nix} and \cite{van2005continuous} developed basic infrastructures which can automatically build, package and deploy individual code changes. These ideas in turn stir more and more advanced environments to offer the ability of running sets of unit, integration or even acceptance tests in a massively parallel way in the shorter time between releases (\cite{downer2012some}). \cite{humble2010continuous} posed the concept of continuous delivery, which uses highly automated infrastructure to deploy new releases in record time. Amazon, for example, deploys on average every 11.6 seconds (\cite{jenkins2011velocity}), achieving more than 1,000 deployments per hour.

Despite all the work discussed above on pushing for shorter release cycles, empirical evidence that directly shows its improvements on product quality is lacked. \cite{khomh2012faster} report that bugs are fixed faster under rapid release models, but proportionally less bugs are fixed compared to the traditional release model based on their study of release model of Firefox. \cite{marschall2007transforming} found that short release cycles require a more steady flow of releases to control the number of reported bugs. Releasing too frequently not only decreases the time to run tests, but also might make customers weary of yet another update to install (\cite{porter2006techniques}, \cite{jansen2006ten}), which is in a way the same situation in app stores.

\subsection{Release Engineering for Mobile Apps} 
Considering app stores, Martin et al. ~\cite{martin2015app} introduced the concept of App Sampling Problem, a threat to validity for app store analysis due to the inherent biases associated with partial information. This threat can be ameliorated by using inferential statistical techniques and/or by carefully defining the scope of claims of apps. In this paper, we alleviate the App Sampling Problem by limiting our claims to the most consistently popular apps in each category of Google Play and the `significant updates'.

There has been considerable work that studies app releases which can be divided into three aspects:

Concerning about the general update behavior of app, \cite{heitkotter2012evaluating} investigated release times and user activity in the Apple App Store, and concluded that Sunday evening was the best time for deploying games in 2011. Their study also found that version updates were an effective strategy for raising an app's rank in the store. In 2013 \cite{datta2013app} studied review counts from the Apple App Store, and found that apps received more reviews after deploying updates on Thursdays or late in the week. \cite{mcilroy2016fresh} found that only {1\%} of the studied apps received more than one update per week, and only {14\%} were updated in a two-week period by studying update frequencies in the Google Play store after mining data of 10,713 mobile apps, the authors also indicated that rating was not affected by update frequency, which is not the same in our paper.

The second aspect is about the update purpose and its general effect with different app release patterns. \cite{lee2014determinants} studied the factors that affect an app’s likelihood of staying in the top charts in the Apple App Store in 2014 . They found that free apps are more likely to `survive', and that frequent feature updates are the most important factor in ensuring their survival, along with releasing in smaller categories. In 2015 Gui et al. found from 21 apps from Google Play with frequent releases, that {23\%} of their releases contained ad-related changes (\cite{Gui2015Truth}). \cite{comino2016updates} studied the top 1,000 apps in Apple App Store and Google Play, finding that for iTunes, numbers of app releases were more likely to increase when the app performs badly, and that releases could boost downloads. 

The last is about the causal effect of app release. \cite{martin2015causal} conducted a longitudinal study on 1,033 apps mined from Google Play and Windows Phone Store over a 12 month time period. The authors used causal inference to identify the releases with most impact on ratings and downloads. However, the releases actually used in the causal effect analysis are limited to those who have no releases within 3 weeks before or after them. Most of the studies use correlation analysis. For example, the findings of \cite{guerrouj2015influence} indicate that high code churn in releases correlates with lower ratings.

All of these previous findings on app releases tantalizingly aim at the possibility that whether certain releases may have higher impact than others. In fact, the situation of developers is totally a different story. When a new version of an app is about to release, the current condition of the app is known, and the last thing for the developers to decide is when to perform the release considering the update purpose. 

\section{Conclusion}\label{conclusion}
In this work we propose an approach to optimizing release strategy for apps in Google Play. In particular, we use update interval instead of average frequency to characterize release patterns and find influential factors using 41,028 Google Play app releases during 105 days, for all apps that appear on the top lists of 33 categories. For these apps, we found that the update intervals are informative to reveal release patterns. App ranking, rating trend, and update purpose all play a role in affecting the update results. High-ranked apps tend to achieve higher ratings after a successive update while low-ranked apps tend to achieve lower ratings; a successive update can be helpful to reverse a descending trend of app rating; bug fixing is more welcomed in successive updates while feature changing have more positive feedback in sparse updates. We have shown that a Multinomial Naive Bayes model can well describe the update effects and the factors. Optimized update interval from this model suggests developers to improve their release strategy. 
In the future we plan to combine our previous work on user management analysis (\cite{IMC15Li, WWW15Li, ICSE16Lu, TOIS17Liu, TSE17Liu}) with this work to develop the release-pattern analytic tool, and provide it as a web service (\cite{SOSE13Ma, Splash12Wang, Chinaf14Liu, ICWS15Liu, Liu2009Discovering, ICSW15Ma}).

\bibliographystyle{ACM-Reference-Format}
\bibliography{sigproc}

\end{document}